%% file: UserGuide.tex
\documentclass[12pt,a4paper]{article}
\usepackage[cp1251]{inputenc}
\usepackage[english]{babel}
\usepackage[pdftex]{graphicx}
\usepackage{amsmath,amssymb,epsfig,amsfonts,multicol,floatflt}
\usepackage{wasysym}
\usepackage{makeidx}
\usepackage{longtable}
\usepackage[pdftex, dvipsnames, usenames]{color} 
\usepackage[unicode, bookmarksnumbered]{hyperref} 
\hypersetup{pdfstartview = FitH} 
\hypersetup{pagebackref=true}
\hypersetup{colorlinks} 
\hypersetup{citecolor = Black} 
\hypersetup{linkcolor = Blue} 
\hypersetup{urlcolor = Sepia} 
\hypersetup{bookmarksopenlevel = 3} 
\usepackage{myStyle}

\makeindex

\begin{document}
\begin{titlepage}
\parskip=0mm
\begin{center}
\normalsize This document can be cited as: Choliy, V. 2013,\\
``DDscat.C++ 7.3.0 User and programmer guide''\\ 
http://arxiv.org/abs/xxxx.xxxx\\ 
\end{center}

\vskip2.0cm
\begin{center} 
\bfseries

\LARGE{\bf DDscat.C++ 7.3.0\\}
\vspace{0.5cm}
\LARGE{\bf User and programmer guide}
\vskip1.5cm
\normalsize
{\rm Vasyl Ya.Choliy\\
Taras Shevchenko National University of Kyiv\\}
{\tt Choliy.Vasyl@gmail.com\\}
\vskip1.0cm
{\rm Last revised 24 June 2013}
\end{center}
\vskip2.0cm
{\center {\bf Abstract\\}}

\ddscatcpp 7.3.0 is a freely available open-source C++ software package app\-ly\-ing the ``discrete
dipole approximation'' (DDA) to calculate scattering and absorption of electromagnetic waves by targets
with arbitrary geometries and a complex refractive index. \ddscatcpp is a clone of well known 
\ddscat Fortran-90 software. We refer to \ddscat as to the parent code in this document. 
Versions 7.3.0 of both codes have the identical functionality but the quite different implementation. 
Started as a teaching project, the \ddscatcpp code differs from the parent code \ddscat in programming  
techniques and features, essential for C++ but quite seldom in Fortran.

As \ddscatcpp in its current version is just a clone, usage of \ddscatcpp for electromagnetic 
calculations is the same as of \ddscat. Please, refer to ``User Guide for the Discrete Dipole
Approximation Code DDSCAT 7.3'' \cite{UGFortran} to start using the code(s).

This document consists of two parts. In the first part we present Quick start guide for users who want to
begin to use the code. Only differencies between \ddscatcpp and \ddscat are explained. 
That is why a lot of references to \cite{UGFortran} are in the first part. 
The second part of the document explains programming tips for the persons who want to change the code, to add the
functionality or help the author with code refactoring and debugging.

The author is grateful to thanks B.Draine and P.Flatau for positive and warm attitude to our efforts 
and the permission to use the name \ddscatcpp for the new code.
 
\vskip1.0cm
\vspace{2cm}
\end{titlepage}

\fontsize{12.4pt}{16.2pt} \selectfont

\tableofcontents\newpage

\section{Introduction}
\label{Section::Introduction}

Electromagnetic energy is scattered or absorbed by targets. It is an isolated grain (of arbitrary geometry
and possibly with complex refractivity index) or 1-d or 2-d periodic structure of unit cells.
According to discrete dipole approximation (DDA), the target is approximated with the array of polarizable
particles (dipoles). The current version of the code works with electric dipoles only. Adding the magnetic properties 
is one of our next steps. 

The theory of DDA and explanation of \ddscat algorithms are given in \cite{UGFortran} and in references therein. 
The current version of \ddscat is 7.3.0 and we refer here this code as the parent one. The parent code User guide 
\cite{UGFortran} is an appropriate and necessary book to start using the code.

Our code \ddscatcpp is the \ddscat rewritten in C++. Current version of \ddscatcpp is a clone
of the parent code but it contains some C++ specific features to make it easily modifiable and portable.
We plan to extend the \ddscatcpp functions in the future.
It is open source and freely downloadable from \verb^http://code.google.com/p/ddscatcpp/^. 
At the beginning of the story the idea was to have a good software for the students to study the photonics and IT 
in the single package. Step-by-step the code has changed and now we have the code with another 
design and architecture but mostly with the same functionality. 

Like the parent code, \ddscatcpp might be usable for many applications without modification.
Anyway the users are encouraged to experiment with the code.
Cloned C++ version, like the parent code, is distributed in the hope that the code will be useful.
 
If you publish results obtained using \ddscatcpp, please:
\begin{itemize}
\item acknowledge the source of the code, make the citation of the paper \cite{MyAASP} and mention the parent code
by referencing the articles of B.Draine and P.Flatau \cite{Draine1} -- \cite{Draine5},
\item comply with the GNU General Public License: you may copy, distribute, and/or modify the 
software identified as coming under this agreement. If you distribute copies of this software, 
you must give the recipients all the rights which you have. 
\end{itemize}

The author \verb^Choliy.Vasyl@gmail.com^ will be glad to read the mail from new \ddscatcpp users or collaborators with  
your statement that you are new user. It will help in communication.
Your reports of the bugs and errors in the code (if any) and your recommendations to the author are welcome.

The code and related data is collected here: \verb^http://code.google.com/p/ddscatcpp/^. 
Please, visit it from time to time: we have a plan to keep the site up to current status of the code.
We will keep some small bug-base there too.

\newpage
\section{Quick start}
\label{Section::QuickStart}

\subsection{Downloading the source code and examples}
\label{Subsection::QuickStart::Download}

\ddscatcpp is written in standard C++ with a little usage of STL. The style is quite plain and there are no
extreme C++ features used. This means that the code should be portable to any platform with C++ compiler
installed. We tested the distribution on Linux (Debian, Ubuntu), MacOS 10.5.8 where gcc is the preferable compiler, 
and on Windows with Microsoft VC 7.1 and Intel C++ 11 compilers. The code was tested on all those platforms 
with Qt 4.7.4 (it again uses gcc).

Normal (ordinal) makefiles are included in the distribution together with Microsoft VC 7.1, Qt 4.7.4 and Xcode projects.
Please find your preferable (appropriate) IDE project and use it to compile the code.

The delivery consists of the single file \verb^ddscatcpp.7.3.0.zip^. As Google does not provide Downloads,
but only SVN or Git access to the code, please, find delivery at \verb^space.univ.kiev.ua/Choliy/DDscatcpp/^
or use SVN from Google code site.

The user may use precompiled binary files. A lot of them are presented in download section of the \ddscatcpp Google code site.

\subsection{Delivery overview}
\label{Subsection::QuickStart::DeliveryOverview}
Select the empty directory, for example \verb^ddscatcpp^, and unpack the delivery package there. 

Explanation of the top level directory tree is given below:
\begin{description}
\item[.:] project files for MSVC and the main makefile;
\item[Bin:] to store binary files generated with MSVC;
\item[BinA:] to store binary files generated with ordinal makefiles;
\item[BinQt:] to store binary files generated with Qt at any platform;
\item[BinX:] to store binary files generated with Xcode at Mac;
\item[CallTarget:] source files of CallTarget program;
\item[CallTarget2:] empty, reserved for Python code of CallTarget2 program;
\item[DDscat:] source files of \ddscatcpp;
\item[DDscatQt:] project files for Qt at any platform;
\item[DDscatX:] project files for Xcode at Mac;
\item[diel:] dielectric (and in the future - magnetic) property files;
\item[Doc:] this User guide;
\item[Fourierlib:] source files of FFT library;
\item[General:] source files of general kind;
\item[Postprocess:] source files of DDpostprocess program;
\item[Processlib:] source files of the library for Postprocess, Readnf1, Readnf2;
\item[Readnf1:] source files of Readnf1 program;
\item[Readnf2:] source files of Readnf2 program;
\item[Results:] scripts and parameter files for an extra examples, not included in the parent code distribution;
\item[Solverlib:] source files of the library of Conjugated Gradient solvers;
\item[Targetlib:] source files of Target library;
\item[TestDDscat:] testing subsystem (a little outdated);
\item[Tests:] scripts and parameter files to run parent code examples;
\item[VtrConvert:] source files of VtrConvert program (a little outdated);
\item[Vtrlib:] source files of the Vtr library;
\item[Xml:] Xml related stuff.
\end{description}

\subsection{Compiling the code}
\label{Subsection::QuickStart::Compiling}
The compilation of the code is essential for any platform. Open appropriate project file and select 
Build All from the main menu of your IDE. We made serious efforts to make the code compilable without 
warnings. Warnings during the compilation of the code should be interpreted as if something went wrong.

\subsubsection{MSVC}
The solution file \verb^DDscat.sln^ resides in main directory. It was created with MSVC 7.1. 
Select Build -- Rebuild All from the main menu. Compilation results will be collected in \verb^Bin^ directory.
Binary xml distribution files should be copied in \verb^Bin^ before build.

\subsubsection{Qt}
The main project file \verb^DDscatQt.pro^ resides in \verb^DDscatQt^ directory. It was created with Qt Creator 2.6.1 
and Qt 4.7.4. Select Build -- Rebuild All from the main menu. Compilation results will be collected in \verb^BinQt^ directory.
Binary xml distribution files should be copied in \verb^BinQt^ before build.

The user can use Qt project to build \ddscatcpp under Linux or Mac OS X. Some additional work is necessary 
to replace pathes and library extensions in \verb^*.pro^ files.

\subsubsection{Mac OS}
For Xcode users we provide Xcode project files in \verb^DDscatX^ directory. It was created with Xcode 3.0 at Mac OS 10.5.8.
The main project file is \verb^DDscat.xcodeproj^. Select Rebuild from main Xcode directory. Compilation results will be collected 
in \verb^BinX^ directory. Binary xml distribution for Mac OS X 10.5.8 is already installed as a system component. 
Anyway, the project is configured as if binary xml files are already present in \verb^BinX^ before build.

\subsubsection{Makefiles}
For Linux users it is essential to have the distribution based upon the autoconf tool. 
Despite of that we provide just an ordinal makefiles, stored in every directory of the distribution.
Autoconf is planned for the future releases.
Just type \verb^make^ in a console opened at the main directory location and gcc (if exists) will produce the distribution in 
\verb^BinA^ directory. Binary xml distribution files should be copied in \verb^BinA^ before make.

\subsubsection{Precompiled binaries}
The user can download all binaries stored in a single \verb^zip^ file from authors web site
\verb^space.univ.kiev.ua/Choliy/DDscatcpp/^. The files contain all necessary binaries including xml libraries. 
These files are copies of \verb^Bin*^ directories. Filenames are \verb^DDscatcpp.7.3.0.MSVC.zip^, \verb^DDscatcpp.7.3.0.WinQt.zip^, 
\verb^DDscatcpp.7.3.0.Xcode.zip^. 

\section{Running the application}
\label{Section::Running}

\subsection{Sequential version}
After the successfull compilation go to an appropriate \verb^Bin^ directory where all necessary binary 
files should be already collected bu build process. The user should identify \verb^DDscat^, \verb^Readnf1^, \verb^Readnf2^, 
\verb^CallTarget^, \verb^VtrConvert^, \verb^DDpostprocess^ executive files and \verb^Fourierlib^, \verb^Solverlib^, 
\verb^Targetlib^, \verb^Vtrlib^ libraries and third party libraries \verb^libXml2^, \verb^iconv^, \verb^zlib1^.

\ddscatcpp may be run with the single parameter (the name of par file) or without parameters. In the latter case
the \ddscatcpp executive search the current directory for \verb^ddscat.par^ file. If the file is not found, the executive
will search for \verb^ddscatpar.xml^ and use it.

\ddscatcpp generates a lot of messages in \verb^stdout^ (file for messages, normally attached to display). We 
recommend to redirect \verb^stdout^ to some log file to analyse it in post-run mode. Error messages are written into
\verb^stderr^ (error file, normally attached to display too) which is better to leave on screen.

\subsection{Parallel version}
All the parallel code in \ddscatcpp 7.3.0 is temporary disabled.
MPI and OpenMP codes will stay disabled until version 7.3.2 while the CUDA - based code will be released in 7.3.1 
after all tests. MPI and OpenMP codes are present in C++ code but are absolutely not tested.
User might catch unpredictable errors (as a minimum, a lot of compiler messages) if tries to 
use MPI or OpenMP with current version of \ddscatcpp.

\section{Parameter file}
\label{Section::ParameterFile}
The \ddscatcpp may be controlled with parameters file of the parent code but some additional freedom
in the parameter file is allowed. There are two special compositions: \verb^Water^ and \verb^Ice^ allowed 
as the composition file names. These are reimplementation of the \verb^refwat^ and \verb^refice^ routines
from \ddscat 6.

\subsection{Text parameter file}
All string parameters may be presented without putting into apostrophes (if they do not contain blank characters). 
So, \verb^’GPFAFT’^ like in the parent code and \verb^GPFAFT^ are identical and are allowed. 
All the lines starting with a) the apostrophe and the blank or b) the apostrophe followed with three asterisks or 
c) the exclamation sign are interpreted as an comments and are just skipped. The user may add as many of such lines 
as he need, for example for documenting reasons.

The target name may be any length single word with free capitalization and all underscore symbols ignored 
by \ddscatcpp. That is why \verb^SPH_ANI_N^ and \verb^SphaniN^ or even \verb^S__p_HAn___iN__^ are identical  
and are allowed.

\ddscatcpp makes memory allocation only once during the target loading. That is why 8th and 9th lines of
the parameter file are ignored but should be present in the file.

In the definition of composition files after line 13 there might be a lot of file names given in the parameter
file. \ddscatcpp allows the usage of equality sign after some amount of composition files given. It means that
all already given file names will be cyclically repeated until their amount become equal to \verb^NCOMP^.

For example there is a portion of a normal par file:
\begin{verbatim}
>>>> begin
12         = NCOMP = number of dielectric materials
`../diel/m1.33_0.01' = file with refractive index 1
`../diel/m1.50_0.01' = file with refractive index 2
`../diel/m1.50_0.02' = file with refractive index 3
=
>>>> end
\end{verbatim}
This means that there are 12 composition files in the example and composition 4 is equal to composition 1, composition 5 to 2,
and so on until composition 12 (obviously equal to 9, then equal to 6 then equal to 3). The error message is generated when 
amount of the given file names exceeds \verb^NCOMP^.

\subsection{Xml parameters file}
\ddscatcpp can be controlled with xml parameter files. DTD for xml parameter files \verb^DDscatcpp.dtd^ 
resides in \verb^Xml^ subdirectory. An example \verb^xml^ parameter file for \verb^Rctglprsm^ test from \ddscat is given 
below. Line numbers the every line starts with are not the part of an \verb^Xml^ file and are present for the orientation.
Users are allowed to add any amount of comment lines in the \verb^Xml^ parameter file. Just start them with \verb^<!--^ and 
end with \verb^-->^, see lines 11 and 12 for example, so multiline comments are allowed. 

\begin{verbatim}
  1   <?xml version="1.0" encoding="utf-8"?>
  2   <!DOCTYPE DDScatParameterFile SYSTEM "ddscatcpp.dtd">
  3   <DDScatParameterFile ver="7.3">
  4   <Preliminaries>
  5      <Cmtorq Value="NOTORQ"/>
  6      <Cmdsol Value="PBCGS2"/>
  7      <CmdFFT Value="GPFAFT"/>
  8      <Calpha Value="GKDLDR"/>
  9      <Cbinflag Value="NOTBIN"/>
 10   </Preliminaries>
 11   <!--NCOMP number of dielectric materials
 12       DIELEC file with refractive index 1-->
 13   <TargetGeometryAndComposition>
 14      <Cshape Name='Rctglprsm'/>
 15      <Shpar Pos="1" Value="16" Comment="x size of the target"/>		
 16      <Shpar Pos="2" Value="32"/>
 17      <Shpar Pos="3" Value="32"/>
 18      <Ncomp Amount='1'>
 19         <Dielec Pos="1" File="../diel/Au_evap"/>
 20      </Ncomp> 
 21   </TargetGeometryAndComposition>
 22   <NearfieldCalculation Nrfld="0">
 23      <Extendxyz Xm="0.0" Xp="0.0" Ym="0.0" Yp="0.0" Zm="0.0" Zp="0.0"/>
 24   </NearfieldCalculation>
 25   <Tol Value="1.00e-5"/>
 26   <Mxiter Value="300"/>
 27   <Gamma Value="1.00e-2"/>
 28   <Etasca Value="0.5"/>
 29   <VacuumWavelengths First="0.5000" Last="0.5000" HowMany="1" How="LIN"/>
 30   <Nambient Value="1.000"/>
 31   <Aeff First="0.246186" Last="0.246186" HowMany="1" How="LIN"/>
 32   <IncidentPolarization Iorth="2">
 33      <PolarizationState>
 34         <X Re="0" Im="0"/>
 35         <Y Re="1" Im="0"/>
 36         <Z Re="0" Im="0"/>
 37      </PolarizationState>
 38   </IncidentPolarization>
 39   <Iwrksc Value="1"/>
 40   <PrescribeTargetRotations>
 41      <Beta  Min="0." Max="0." Number="1"/>
 42      <Theta Min="0." Max="0." Number="1"/>
 43      <Phi   Min="0." Max="0." Number="1"/>
 44   </PrescribeTargetRotations>
 45   <SpecifyFirst Iwav="0" Irad="0" Iori="0"/>	
 46   <S_ijMatrix Number="6">
 47      <ij Value="11 12 21 22 31 41"/>
 48   </S_ijMatrix>
 49   <ScatteredDirections Cmdfrm="LFRAME" Nplanes="2">
 50      <Plane N="1" phi="0."  MinThetan="0." MaxThetan="180." Dtheta="5"/>
 51      <Plane N="2" phi="90." MinThetan="0." MaxThetan="180." Dtheta="5"/>
 52   </ScatteredDirections>
 53   </DDScatParameterFile>
\end{verbatim}
There should not be difficulties in understanding of the \verb^xml^ parameter file. Most of its lines are 
quite self-explanatory and are easily mapped onto the lines of the text par file. Presented example contains all 
possible tags. Sometimes, when there is no necessity, some of the tags may be dropped. For example if the user 
does not plan to do nearfield calculations one can omit lines 22 - 24, or 27th if there is no gamma used in 
calculations, etc. Comment attribute in \verb^Shpar^ element may be dropped too.

\section{Test package}
\label{Section::TestPackage}
The \verb^Tests^ and \verb^Results^ directories contain scripts and parameter files for all parent code
examples and all targets mentioned in \ddscat User guide \cite{UGFortran}. The \verb^Tests^ parameter 
files are identical to those of \ddscat User guide, but \verb^Results^ ones are quite artificial and should
be used only for illustration. Main difference between \verb^Tests^ and \verb^Results^ is that all results do nearfield
calculation and have MayaVi2 snapshots.

Any test or result resides in its own direstory. The directory contains par and xml parameter files together with 
target explanation files (targ) if any.

To run tests or results go into appropriate directory and run \verb^RunAll*^ script.
The script run \ddscatcpp for all subdirectories and then run \verb^Readnf1^ and \verb^Readnf2^ for them to produce
\verb^vtr^ files for \verb^MayaVi2^ and field crossing along the line. All scripts are quite elementary.

To add new Result to the \verb^Result^ directory one should:
\begin{itemize}
\item create the own directory \verb^DirectoryName^,
\item copy \verb^RunResult.bat^, \verb^RunReadnf1.bat^ and \verb^RunReadnf2.bat^ into it from any subdirectory, 
copy extra files, like target explanation file into it,
\item modify them if you need some extra files, 
\item run \verb^TheResult DirectoryName^ script,
\item run \verb^TheReadnf1 DirectoryName^ script if you need near target field,
\item run \verb^TheReadnf2 DirectoryName^ script if you need field crossing along the line,
\item find resulting files and logs in \verb^DirectoryName^.
\end{itemize}

Please, refer to \ddscat User guide \cite{UGFortran} for target explanations. 
The only targets explained here are new ones. In \verb^Tests^ and \verb^Results^ directories one can find
all necessary things to run the tests and results including the MayaVi2 snapshot from our runs.

\subsection{EllipsoN: N aligned homogenous isotropic ellipsoids}
The target consists of N ellipsoids identical in size but possibly different in composition placed along x axis.
There are 5 parameters:
\begin{itemize}
\item \verb^SHPAR_1^ = length of the ellipsoid in x direction;
\item \verb^SHPAR_2^ = length of the ellipsoid in y direction;
\item \verb^SHPAR_3^ = length of the ellipsoid in z direction;
\item \verb^SHPAR_4^ = N - number of ellipsoids;
\item \verb^SHPAR_5^ = distance between ellipsoids surfaces along x direction.
\end{itemize}
User should provide N compositions and set \verb^NCOMP = N^.

\begin{figure}
\centering{\includegraphics[width=140mm]{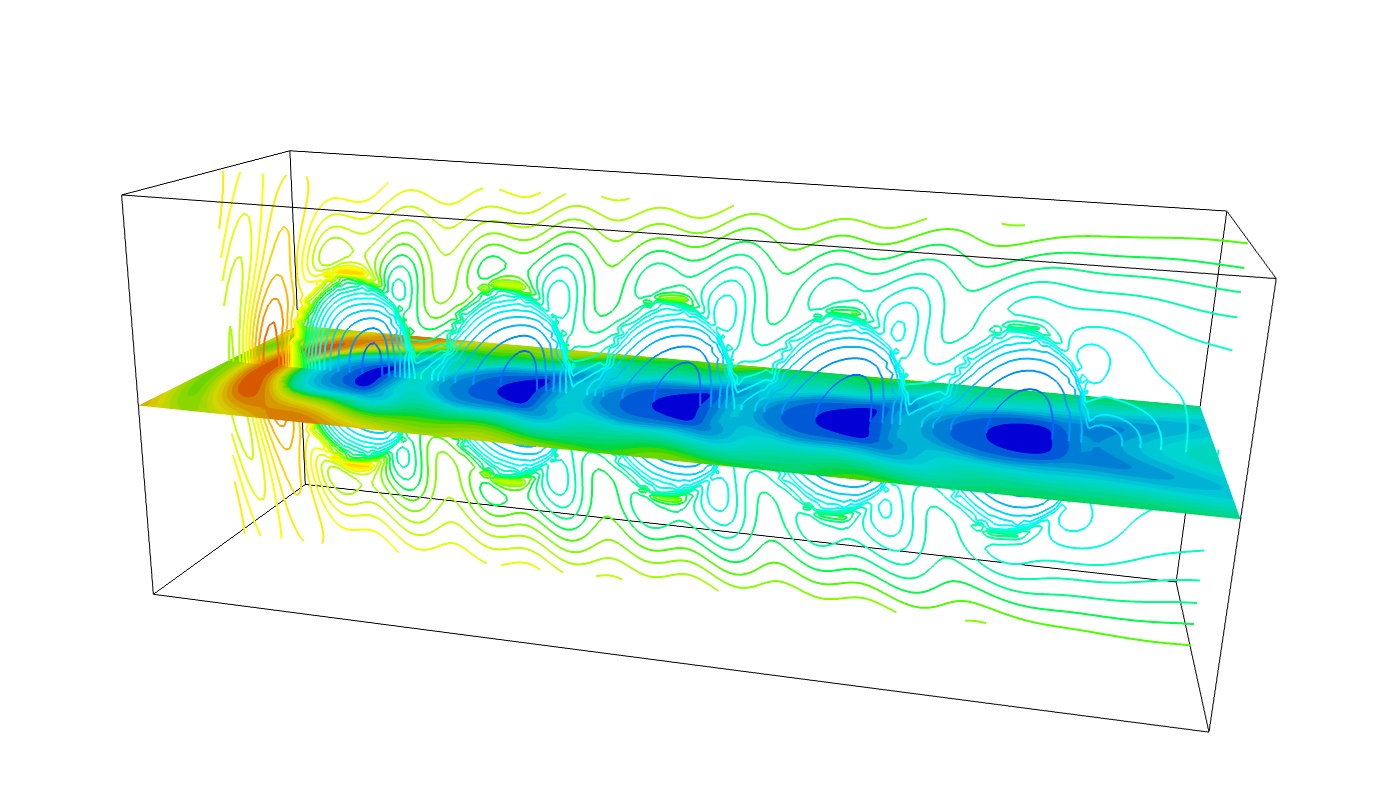}}
\caption{General view of the electric field in EllipsoN result.}
\label{Figure::Ellipson}
\end{figure}

\noindent The portion of example calculation of the \verb^par^ file is copied just below with MayaVi2 visualization
of the electric field on Fig.~\ref{Figure::Ellipson}:
\begin{verbatim}
'ELLIPSON' = CSHAPE*9 shape directive
24. 36. 30. 5. 6. = shape parameters 1 - 5
5         = NCOMP = number of dielectric materials
'../diel/m0.96_1.01' = file with refractive index 1
'../diel/m0.96_1.01' = file with refractive index 2
'../diel/m0.96_1.01' = file with refractive index 3
'../diel/m0.96_1.01' = file with refractive index 4
'../diel/m0.96_1.01' = file with refractive index 5
'**** Additional Nearfield calculation? ****'
1 = NRFLD (=0 to skip nearfield calc., =1 to calculate nearfield E)
0.1 0.1 0.5 0.5 0.5 0.5 (fract. extens. of calc. vol. in -x,+x,-y,+y,-z,+z)
\end{verbatim}

As we already mentioned above if \verb^par^ file contains a lot of composition files, they can be replaced with equality sign,
that is why the portion of the \verb^par^ file can be replaced with:
\begin{verbatim}
'ELLIPSON' = CSHAPE*9 shape directive
24. 36. 30. 5. 6. = shape parameters 1 - 5
5         = NCOMP = number of dielectric materials
'../diel/m0.96_1.01' = file with refractive index 1
=
'**** Additional Nearfield calculation? ****'
1 = NRFLD (=0 to skip nearfield calc., =1 to calculate nearfield E)
0.1 0.1 0.5 0.5 0.5 0.5 (fract. extens. of calc. vol. in -x,+x,-y,+y,-z,+z)
\end{verbatim}

\subsection{AniEllN: N aligned homogenous anisotropic ellipsoids}
The target consists of N anisotropic ellipsoids identical in size but possibly different in composition placed along x axis.
There are 5 parameters identical to those of \verb^ELLIPSON^.
User should provide 3*N compositions and set \verb^NCOMP = 3*N^.

\begin{figure}
\centering{\includegraphics[width=140mm]{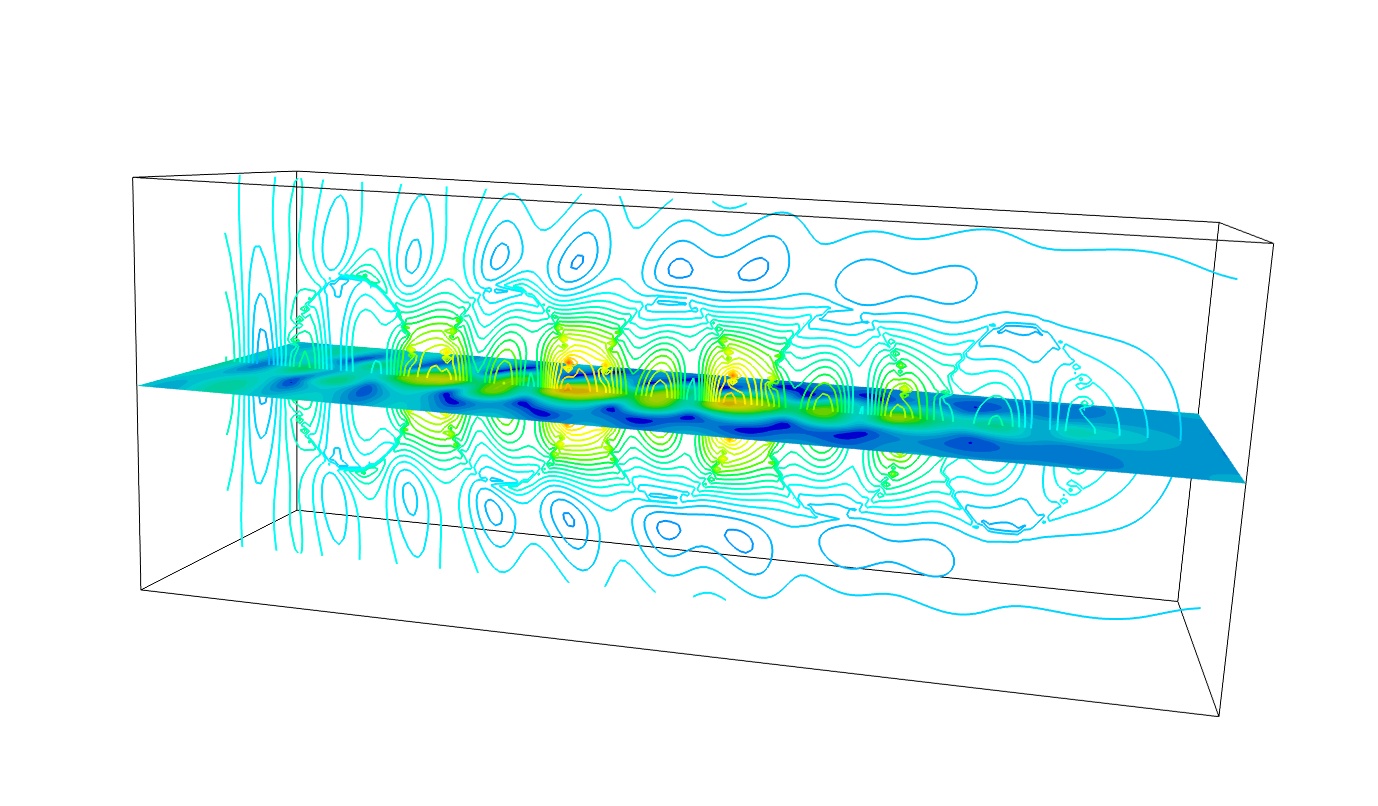}}
\caption{General view of the electric field in AniEllN result.}
\label{Figure::AniEllN}
\end{figure}

\noindent The portion of example calculation of the \verb^par^ file is copied just below with MayaVi2 visualization
of the electric field on Fig.~\ref{Figure::AniEllN}:
\begin{verbatim}
'ANIELLN' = CSHAPE*9 shape directive
24. 36. 30. 5. 6. = shape parameters 1 - 5
15         = NCOMP = number of dielectric materials
'../diel/m1.33_0.01' = file with refractive index 1
'../diel/m1.50_0.01' = file with refractive index 2
'../diel/m1.50_0.02' = file with refractive index 3
=
'**** Additional Nearfield calculation? ****'
1 = NRFLD (=0 to skip nearfield calc., =1 to calculate nearfield E)
0.1 0.1 0.5 0.5 0.5 0.5 (fract. extens. of calc. vol. in -x,+x,-y,+y,-z,+z)
\end{verbatim}

\subsection{OctPrism: octagonal prism}
This target represent the single octagonal prism particle. It is mostly the same as hexagonal prism, but only
one orientation (main prism axis lies along the x coordinate) is provided now.
There are two parameters:
\begin{itemize}
\item \verb^SHPAR_1^ = length of the prism in x direction;
\item \verb^SHPAR_2^ = distance between opposite vertices of one octagonal face.
\end{itemize}
User should provide 1 composition and set \verb^NCOMP = 1^. Target name is \verb^OCTPRISM^.

\begin{figure}
\centering{\includegraphics[width=140mm]{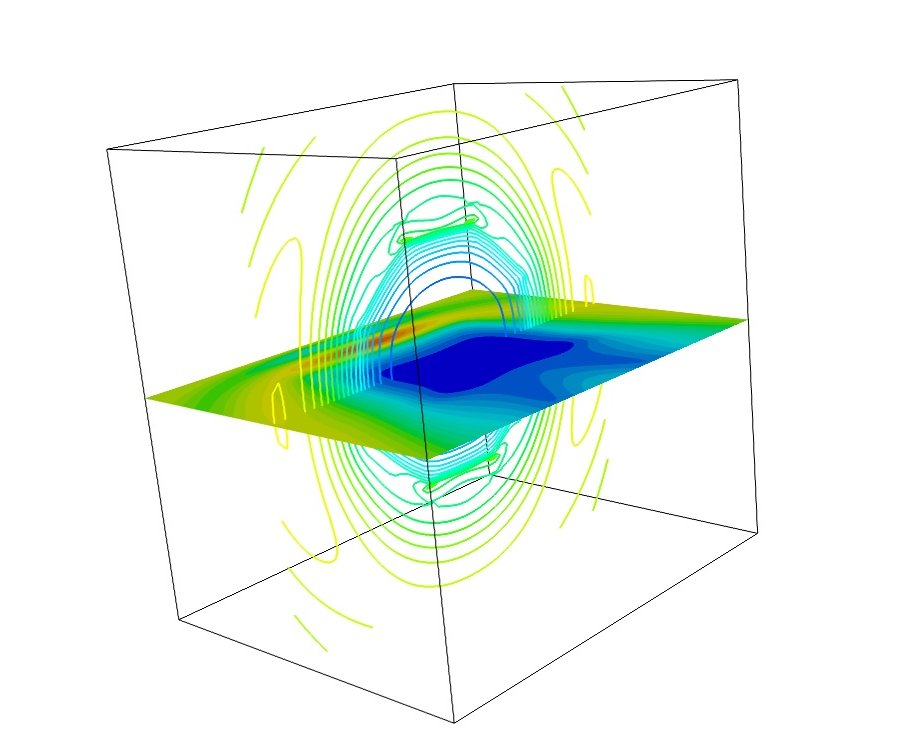}}
\caption{General view of the electric field in OctPrism result.}
\label{Figure::OctPrism}
\end{figure}

Figure~\ref{Figure::OctPrism} represents the electric field near the octagonal prism target with 
parameters equal to 24. and 30.

\subsection{Octahedron: single isotropic octahedron particle}
\begin{figure}
\centering{\includegraphics[width=140mm]{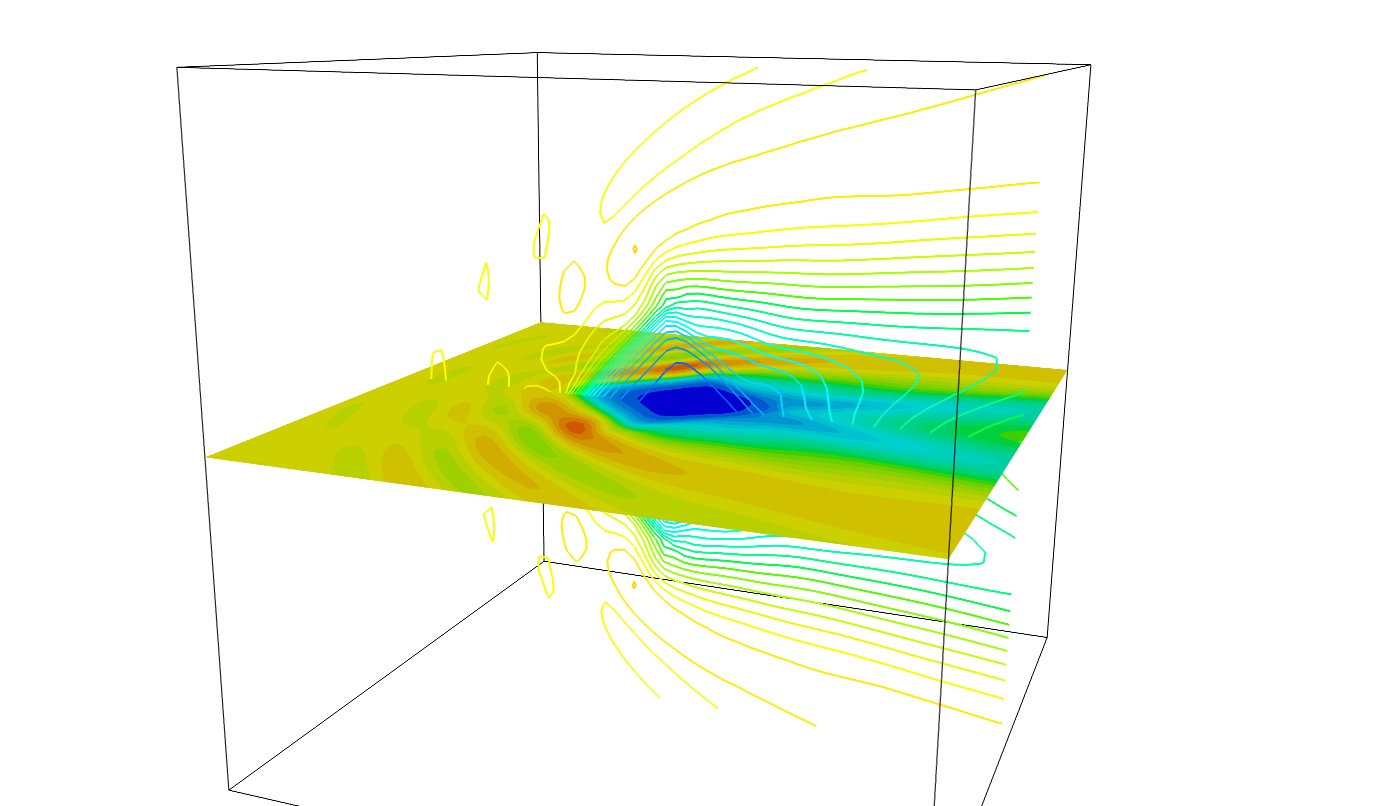}}
\caption{General view of the electric field in Octahedron result.}
\label{Figure::Octahedron}
\end{figure}

The target is octahedron particle. 
Each line that pass through two oposite vertex is parallel to the x,y, and z axis, respectively. 
The target needs the only parameter: a distance between two opposite vertices, or a diameter of escribing sphere.
The target is isotropic, so \verb^NCOMP = 1^ and the user should provide only one composition file.
Figure~\ref{Figure::Octahedron} represents the electric field near the octahedron target with 
parameter equal to 40. Target name is \verb^OCTAHEDRON^.

\subsection{Icosahedron: single isotropic icosahedron particle}
\begin{figure}
\centering{\includegraphics[width=140mm]{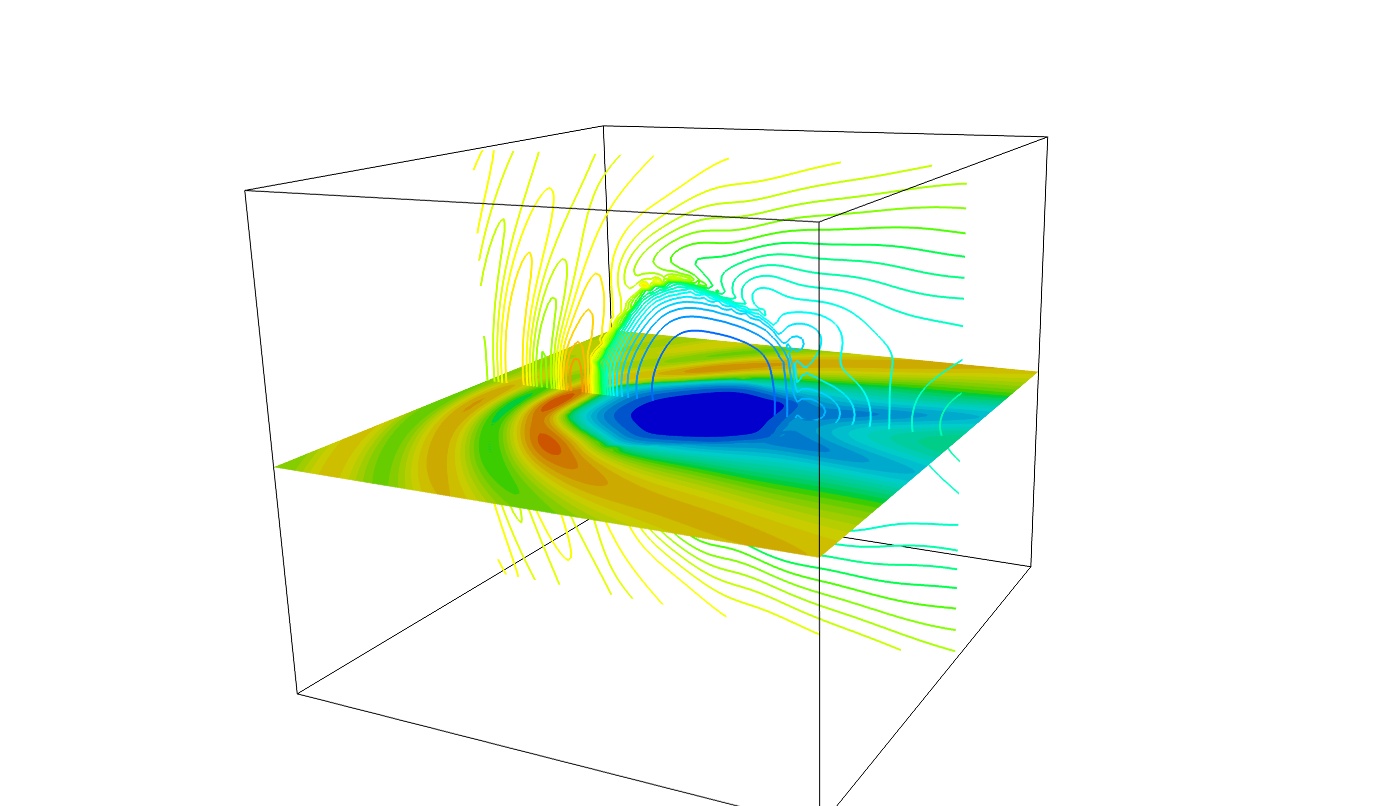}}
\caption{General view of the electric field in Icosahedron result.}
\label{Figure::Icosahedron}
\end{figure}

The target is icosahedron particle oriented like the octahedron one. 
The target needs the only parameter: the lenght of the edge.
The target is isotropic, so \verb^NCOMP = 1,^ and the user should provide only one composition file.
Figure~\ref{Figure::Icosahedron} represents the electric field near the icosahedron target with 
the parameter equal to 24. The target name is \verb^ICOSAHEDRON^.

\subsection{Dodecahedron: single isotropic dodecahedron particle}
\begin{figure}
\centering{\includegraphics[width=140mm]{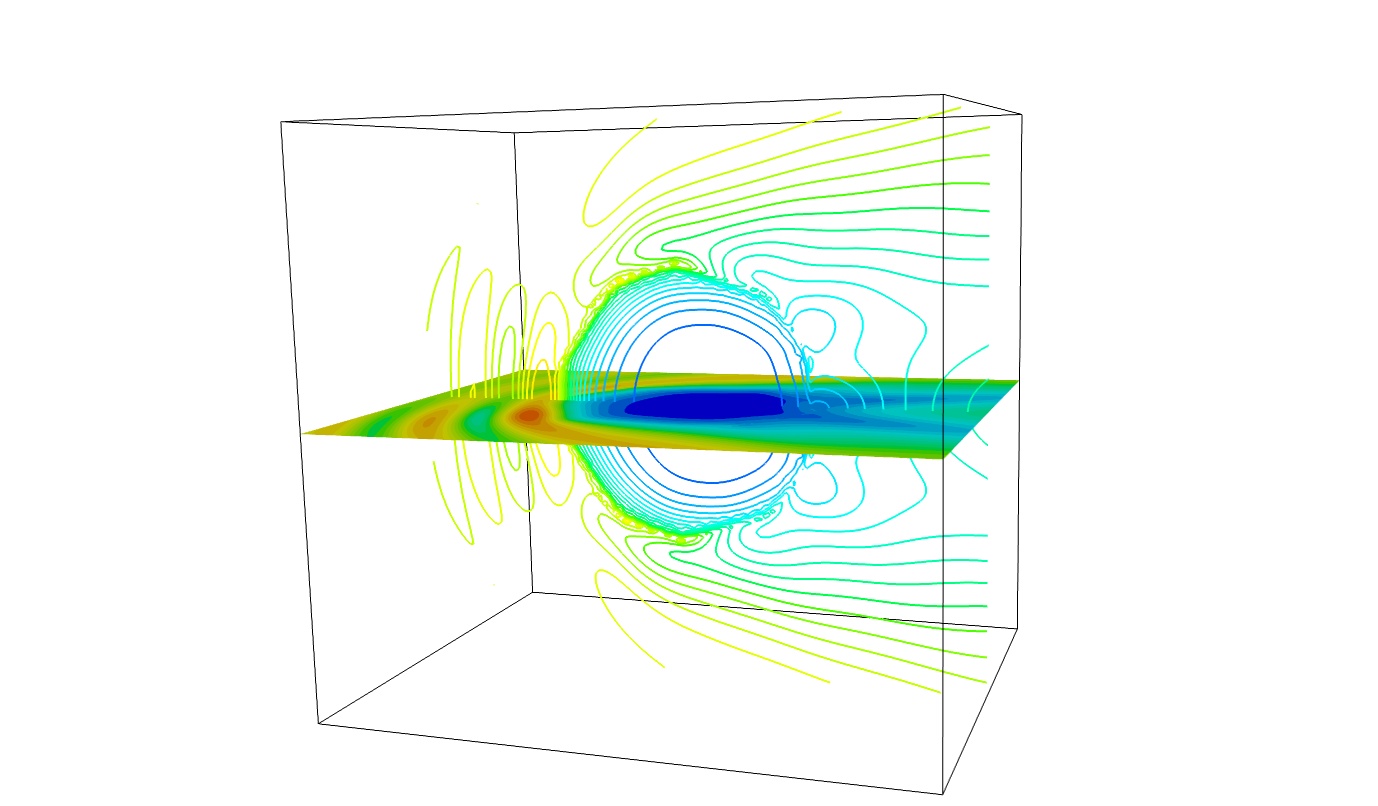}}
\caption{General view of the electric field in Dodecahedron result.}
\label{Figure::Dodecahedron}
\end{figure}

The target is dodecahedron particle. The pentagonal base is parallel to the xy plane and one edge of the base is parallel 
to the y axis. The target needs the only parameter: the lenght of the edge.
The target is isotropic, so \verb^NCOMP = 1,^ and the user should provide only one composition file.
Figure~\ref{Figure::Dodecahedron} represents the electric field near the dodecahedron target with 
the parameter equal to 24. The target name is \verb^DODECAHEDRON^.

\newpage
\section{General programming remarques}
\label{Section::Programming}
The code consists of components with strictly defined communication protocol 
and lightweight replacement, modification and refactoring (plugin paradigm). \ddscatcpp is a set of dynamically linkable libraries with 
defined and fixed communication interfaces.

Examples and testing capabilities are an essential part of the code. As running of all tests consumes a lot of time, we 
don’t use CppUnit but our own code and scripts library to be run on request. All \ddscat tests work fine as a part of \ddscatcpp code.

\begin{figure}[h]
\centering{\includegraphics[width=0.9\columnwidth]{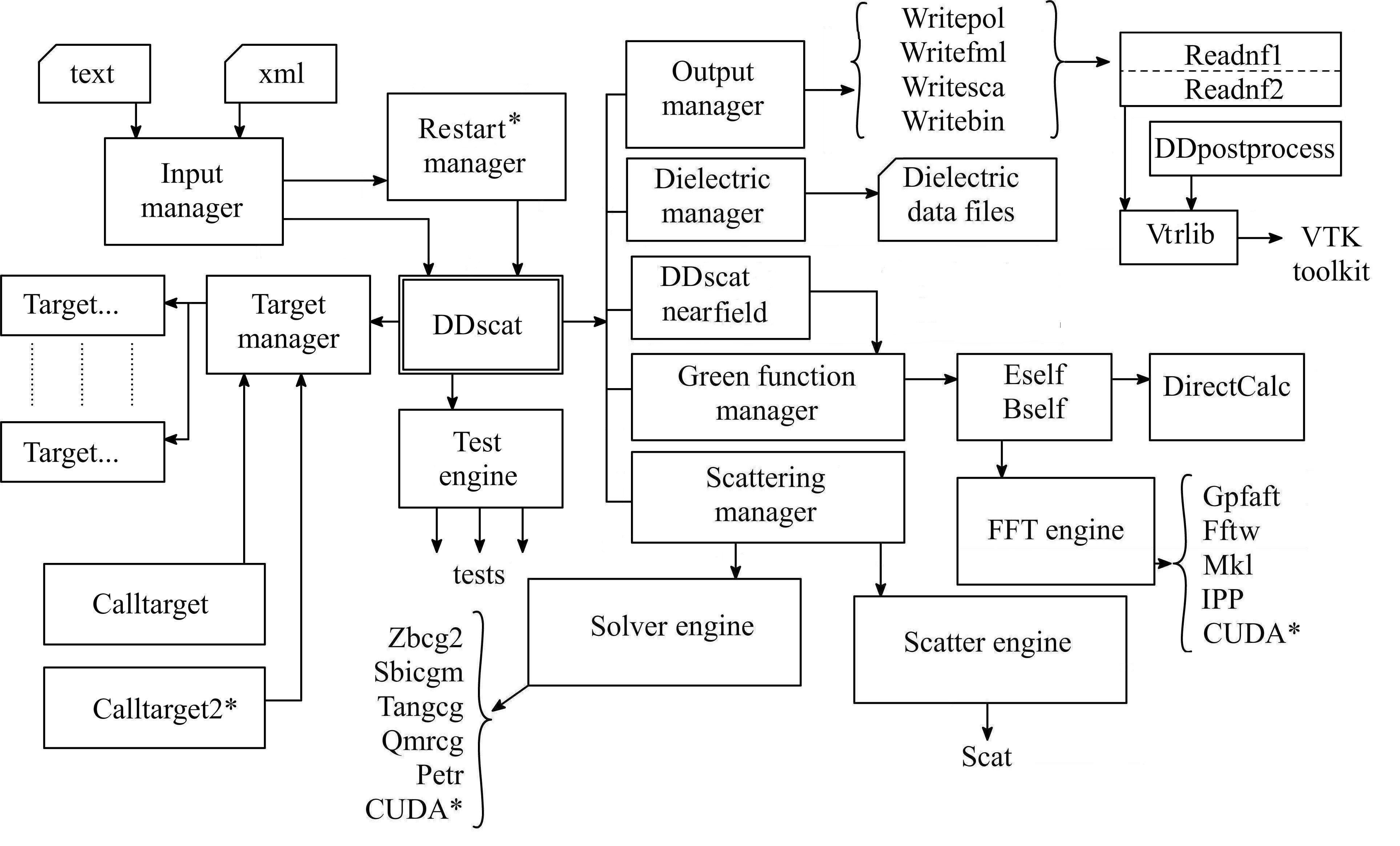}}
\caption{General view of the code architecture.}
\label{Architect}
\end{figure}

The overall view of the architecture and its main blocks are presented on Fig.~\ref{Architect}.
The users familiar with the parent code may easily identify known code blocks. The asterisk as an upper index
marks new code parts, introduced in C++ version. Every code portion is controlled with and is communicated
via the specially designed manager components. These code snippets are singletons.

The code uses C-style of indexing. This means that all indexes in all arrays in the code always starts from zero, not one, 
as it is in fortran.

The code and most of the names are case sensitive, but target names are not. Please, be on guard with it.

The code needs cosmetics and refactoring. Strictly speaking the code is not written in C++, much better to say
that the code is written in C with some amount of classes. That is why the code mostly does not contain STL. It is our 
permanent plan for the future releases: refactoring to add STL and make cosmetics in necessary places but without fanatism.

\section{Target manager}
\label{Section::TargetManager}
Target manager manipulates the targets. The target explains the grains geometry or represents elementary cell to build 1-d 
or 2-d infinite periodic arrays of targets. The parent code contains a lot of different geometries already implemented. 
These are ellipsoids (spheroids), prisms, cylinders, disks, slabs, tetrahedra, possibly with holes and their simple joints. 
Some of the targets are just a combination or multiplications of existing ones. 

\subsection{TargetManager class}
\label{Subsection::TargetManager::TargetManager}
This singleton class controls life cycle of the current target and owns information about it. 
It is build as a class factory with the possibility to self register the target in the factory.

To access and register itself in the factory the target should be accompanied with \verb^REGISTER_TARGET^ macro.
The parameters of the macro are:
\begin{itemize}
\item the name of the target, an ordinal word with any capitalization;
\item the number of target parameters including file name if any;
\item will the target use additional file, false or true boolean value;
\item the position of periods in parameters if the target is periodic, otherwise -1;
\item the number of required composition, 0 - doest'n matter, not 0 - the number of required ncomp in par file;
\item free description of the target in one line.
\end{itemize}

For example, lets register \verb^Beautiful_Particle^ as not periodic target with two parameters including the file name,  
three compositions required. 
The macro for the target is:
\begin{verbatim}
REGISTER_TARGET(Beautiful_Particle,2,true,-1,3,"Flower-like particle")
\end{verbatim}

The macro should be placed somewhere in cpp file of the target class, whose name should be
\verb^Target_Beautiful_Particle^. The macro forces the user to add two mandatory functions to the code, namely
\verb^void Target_Beautiful_Particle::SayHello(FILE *stream)^ and 
\verb^const char *TargetVerboseDescriptor_Beautiful_Particle(int num)^.

The first function is used to write the values of the target parameters into the stream. The second one 
returns string representation of the parameter identified with num value. All of the targets in delivery have 
those functions. Please, use the code for additional information.

\subsection{AbstractTarget class}
\label{Subsection::TargetManager::AbstractTarget}
This class is the core of the TargetManager library. Despite of its name the class does not contain pure virtual 
functions. It is just a good name for the class, capable to represent any possible target.

The class contains:
\begin{itemize}
\item nat - int - the total amount of dipoles (places for dipoles);
\item nat0 - int - the total amount of occupied dipoles;
\item nx, ny, nz - int - the dimension of the target;
\item ixyz[nat0, 3] - int - relative coordinates of occupied dipoles;
\item minJx, maxJx, minJy, maxJy, minJz, maxJz - int - min and max limits of ixyz values;
\item icomp[nat, 3] - short - a dipole composition;
\item ncomp - int - the number of compositions;
\item iocc[nat] - bool - the dipole occupancy sign;
\item shpar - real - target parameters;
\item pyd, pzd - real - periods if the target is periodic;
\item ianiso - the isotropic flag;
\item a1, a2, dx, x0 - known vectors.
\end{itemize}

The most valuable method of AbstractTarget in Build, which just calls a lot of another methods:
\begin{verbatim}
void AbstractTarget::Build(void)
{
    Sizer();                // determines the nx,ny,nz and min/max values of ixyz
    Descriptor();           // creates target descriptor string
    Allocator();            // allocates ixyz, iocc, icomp
    Vector();               // defines a1 and a2 vectors
    VectorX();              // defines x vector
    Printer();              // prints target head and creates target.out file
    ShiftDipolesAndX();     // shift x and ixyz to have the first dipole at (1,1,1)	
    PrepareIaniso();        // recognizes if the target is anisotropic or not
    PreparePyzd();          // prepares periodicities if necesary
}
\end{verbatim}
All the functions from \verb^Sizer^ to \verb^PreparePyzd^ are dummy in AbstractTarget class. The user 
should create its own versions if added a new target to the TargetManager.

Two of the functions, \verb^Sizer^ and \verb^Allocator^ work together to allocate the memory for the target in 
a single job. \verb^Sizer^ determines min and max values of the dipole coordinates and nx, ny, nz - sizes of
the cubic cell to insert the target in, but it does not do the allocation. In most cases it does a dummy 
allocation - everything is doing as if it is an allocation but the only min/max values are determining.
As if cell dimensions are determined, the \verb^Allocator^ really allocates the memory and put the dipole data in.

The main task of the user when added new targets is to place the dipole x, y, z into ixyz and then using \verb^GetLinearAddress^
to determine the correct place of composition data in the iocc array. GetLinearAddress needs correct min/max values.
{\bf All arrays except specially stated use C-like indexing method.}

\subsection{LoadableTarget class}
\label{Subsection::TargetManager::LoadableTarget}
This class is inherited from AbstractTarget and is designed to represent the target with some data in additional 
(loadable during preparation) targ file. There is only one additional function in LoadableTarget class
\verb^void LoadableTarget::Reader(void)^ which is used to load the targ file and add the information from it to the target.

\subsection{How to add new target}
\label{Subsection::TargetManager::HowToAddTarget}
This is step by step receipt to add a new target to the TargetManager. We will do it using an \verb^OctPrism^ target name 
(but in general sence) as an example. The target name should be unique in the \ddscatcpp scope. 

Every target is represented with its class, but having in mind possible future extensions and modifications of the code 
it is better to add two classes: the generic one to represent the properties of the prisms, and the concrete one to represent the 
properties of the octagonal prism. Namely in the case of prisms, we already have a \verb^Tarhex^ class as a generic one and 
a \verb^Target_HexPrism^ as a concrete one. Obvious future refactoring should result in something like 
\begin{verbatim}
Prism -> Tarhex -> Target_HexPrism
      -> Taroct -> Target_OctPrism
\end{verbatim}
but it needs some additional efforts.

\begin{enumerate}
\item Edit the \verb^Targetlib\TargetDefinitions.h^ file and add a string \verb^TargetType_OctPrism^ somewhere in the 
definition of \verb^enum TargetType^.
\item Add a new generic class \verb^Taroct^ to TargetManager inheriting it from \verb^AbstractTarget^ or from \verb^LoadableTarget^
if the new target will read some extra information from a targ file.
\item Add a new concrete class \verb^Target_OctPrism^ to TargetManager inheriting it from just added generic class \verb^Taroct^.
The concrete class name should always be \verb^Target_^ followed with a preselected target name (\verb^OctPrism^ now).
\item Add a \verb^REGISTER_TARGET^ macro somewhere in a cpp file.
\item Add a \verb^const char *TargetVerboseDescriptor_OctPrism(int num)^ function to a cpp file.
\end{enumerate}

Now the user should implement the functions, called from \verb^AbstractTarget::Build^ (see AbstracTarget class above), 
and Reader if the target is loadable. 
  
\section{Solver engine}
\label{Section::SolverEngine}
Solver is a generic name for the code components designed to solve sets of linear equations. The current version of \ddscatcpp
uses only Conjugated gradient (CG) solvers from CGPACK library of P.Flatau\footnote{http://code.google.com/p/conjugate-gradient-lib/}.
These solvers are well explained in \ddscat User Guide \cite{UGFortran}. For \ddscatcpp it is not mandatory to use CG codes.
The user are allowed to add own solvers.

\subsection{AbstractSolver class}
Singleton AbstractSolver class plays two main roles. It is solver factory and solver manager simultaneously.
The class object controls the lifecycle of the solver and manages the access to the solver in use.

To access and register itself in the factory the solver should be accompanied with a \verb^REGISTER_SOLVER^ macro.
The parameters of the macro are:
\begin{itemize}
\item the name of the class the solver is stored in;
\item the string which represents the solver for the user.
\end{itemize}
The last string to be used for identifying the solver in a par file.

For example, lets register \verb^BeautifulSolver^. The macro for the solver is:
\begin{verbatim}
REGISTER_SOLVER(BeautifulSolver,"BEAUTY")
\end{verbatim}

\subsection{How to add new solver}
This is step by step receipt to add new solver to the Solver engine. We will do it using
Beauty solver name as an example. The solver name should be unique in the \ddscatcpp scope.

\begin{itemize}
\item Edit \verb^enum SolMethod^ in \verb^General\Enumerator.h^ and add \verb^SolMethod_Beauty^ string before \verb^SolMethod_End^;
\item edit functions \verb^SolEnumerator^ in \verb^General\Enumerator.cpp^ and add strings for the new solver;
\item add the new class \verb^Solver_Beauty^ to Solverlib subproject, inheriting it from \verb^AbstractSolver^ class;
\item add a macro \verb^REGISTER_SOLVER^ somewhere in a cpp file.
\end{itemize}

Every solver is represented with its class. The user is allowed to name the class members in any useful way, but there should be:
\begin{itemize} 
\item the solver parameters, they should be set with the special function \verb^SetParameters^ before the call of the solver function;
\item the solver function with two mandatory parameters: the initial guess vector and a right side vector and possibly additional parameters;
in the future refactoring we plan to retain only first two parameters;
\item the external function Matvec which is used by the solver to do matrix-vector multiplications; Matvec is set to the solver 
with the \verb^SetMatvec^ function.
\end{itemize}

\section{Field engines and managers}
\subsection{FFT engine}
\label{Section::Another::FFTEngine}
This engine is the another singleton which is used to manage the access to FFT codes. The current version 
of \ddscatcpp is capable to work only with the Gpfaft code of C.Temperton \cite{Temperton}. Usage of FFTW and Intel 
routines are temporary disabled (to 7.3.1).

\subsection{Green function manager}
\label{Section::Another::GreenFunctionManager}
Green function manager organizes direct calculation of electric and magnetic field with Green function approach.
In current \ddscat there is a lot of code duplicates around \verb^Eself^ and \verb^Bself^ routines, which are 
very similar. In \ddscatcpp there is a \verb^Subsystem^ class which is (or may be) instantiated for electric (Eself) 
and/or magnetic (Bself) calculations. That saves a lot of memory and leeds to greater targets.

Green function manager effectively manipulates with the instances of \verb^Subsystem^ class and replaces the calculations 
with data exchange whenever possible.

\subsection{Dielectric manager}
\label{Section::Another::DielectricManager}
Dielectric manager is a little self-made data management class. It controls dielec files, stores dielectric data, provides 
the access to dielectric values. Our future plans include introducing the magnetic dipoles into calculations.
Dielectric manager is already able to manipulate magnetic data too.

\section{Input and Output managers}
\label{Section::InputOutputManager}
Input manager manages \ddscatcpp parameters, which are organized in the \verb^DDscatParameters^ class. It reads par or xml
files and provides the access to the data stores there.

Output manager is responsible for preparing the output files. It comprises all \verb^Write*^ routines from \ddscat
and manipulates the calculation results to create pretty-looking output.

\section{Restart manager}
\label{Section::RestartManager}
Restart manager is in testing phase and will be released in version 7.3.2. Main task of Restart manager is to do 
a restart of the calculation without loss of the data after sudden events, for example an electricity failure. 

\section{Postprocessing}
\label{Section::Postprocessing}
In the current version (7.3.0) \ddscatcpp includes standalone programs:
\begin{itemize}
\item Readnf1 - to create vtr files for vizualization of the electric field with MayaVi2 software;
\item Readnf2 - to make a field crossing along the specified line;
\item Postprocess - component appeared in \ddscat 7.3.0 which do the jobs of both Readnf's.
\end{itemize}

Readnf1 is controlled with a quite elementary par file:
\begin{verbatim}
'w000r000k000.E1'             = name of file with E stored
'VTRoutput'                   = name of VTR output file
1   = IVTR  (set to 1 or 2 to create VTR output with |E| or |E|^2)
\end{verbatim}
the same is for Readnf2:
\begin{verbatim}
'w000r000k000.E1'            = name of file with E stored
1   = ILINE (set to 1 to evaluate E along a line)
-2.0 0.0 0.0 2.1 0.0 0.0 501  = XA,YA,ZA, XB,YB,ZB (phys units), NAB
\end{verbatim}
The par files of those components are the same as in \ddscat 7.2.2, they are just subdivided into 
two parts. The par file for \verb^Postprocess^ is the same as for Readnf of \ddscat 7.2.2.

\section{Calltarget}
\label{Section::Calltarget}
Function of CallTarget is explained in Fortran version User Guide. CallTarget2 is a new wxPython component 
to be released with \ddscatcpp 7.3.2 and will help users to create the targets interactively.

\newpage
\section{Finale}
\label{Section::Finale}
This User Guide is a subject of permanent changes and will stay improving together with \ddscatcpp code.

The source code and some additional information are available at Google code site:
\verb^http://code.google.com/p/ddscatcpp/^

Post your bugs and suggestions via E-mail to the author \verb^Choliy.Vasyl@gmail.com^. Please, provide your 
E-mail addresses and identify yourself as a user or a tester or a hacker or \dots of the code. As if the author will be
able to inform the engaged persons with news, bug fixes, new releases. 

Extension of the Targetlib with new targets are welcome first.

If you plan to use \ddscatcpp please, cite the article 
\begin{verbatim}
Choliy V. 2013, "The discrete dipole approximation code DDscat.C++: 
features, limitations and plans".
Adv.Astron.Spa.Phys., 3, 66-70
\end{verbatim}
and the articles of B.Draine and P.Flatau, referenced in Finale of \ddscat User Guide \cite{UGFortran}.

The \ddscatcpp author appreciates receiving copies of the articles where \ddscatcpp is mentioned.

Great thanks to B.Draine and P.Flatau for positive and warm attitude to our efforts and the permission to use 
the name \ddscatcpp for the new code.

\printindex

\input{litera}


\end{document}

%% file: litera.tex
\newpage
\renewcommand{\refname}{}